\documentclass[pra,showpacs,twocolumn]{revtex4}

\usepackage[english]{babel}
\usepackage{times}
\usepackage{latexsym}

\usepackage{graphicx} 
\usepackage{epsf} 
\usepackage{subfigure} 
\usepackage{amsmath} 
\usepackage{amssymb} 
\usepackage{amsfonts}
\usepackage{bm}
\usepackage{natbib}
\usepackage{epstopdf}

\def\be{\begin{equation}}
\def\ee{\end{equation}}
\def\bea{\begin{eqnarray}}          
\def\eea{\end{eqnarray}}
\def\bi{\begin{itemize}}
\def\ei{\end{itemize}}
\def\bin{\begin{enumerate}}
\def\ein{\end{enumerate}}

\begin{document}

\title{Mott-Insulator Transition for Ultracold Fermions in Two-Dimensional Optical Lattices}


\author{
N. Goldman
}
\affiliation{Center for Nonlinear Phenomena and Complex Systems - Universit\'e Libre de Bruxelles (U.L.B.), Code Postal 231, Campus Plaine, B-1050 Brussels, Belgium}

\date{\today}

\begin{abstract}
In this work we study ultracold Fermions confined in a two-dimensional optical lattice and we explore the Mott-insulator transition with the Fermi-Hubbard model. On the basis of a mean-field approach, we study the phase diagrams in the presence of a harmonic trapping potential. Local Mott-insulator phases are shown to be generally situated in the center of the trap and correspond to a vanishing variance of the local density. We then study the effects induced by rotation on the Mott-insulator phase transition. In particular, we show that the phase boundary reproduces the edge of the Hofstadter butterfly.

\end{abstract}


\maketitle

\section{Introduction}

Today great efforts are devoted to the investigation of fundamental phenomena using ultracold atoms trapped in optical lattices \cite{zollerann,ouradv}. One of the most outstanding breakthroughs in the field concerns the experimental realization of the bosonic Mott-insulator transition \cite{greiner}. The transition from the superfluid to the Mott-insulator is performed in ultracold bosonic optical lattices by increasing the ratio of the on-site interaction to the tunneling amplitude  \cite{fisher,sheshadri}. When the Mott-insulator phase is reached, the system is characterized by a commensurate filling of the lattice. As predicted by Jaksch \emph{et al.}, this quantum phase transition can be achieved by varying the depth of the optical potential \cite{jaksch}. \\

The fermionic Mott metal-insulator transition (MMIT) has been originally predicted for electronic systems \cite{rama}.  Recently it has been suggested that a fermionic Mott-insulator phase should be observed in one-dimensional optical lattices \cite{rigol, liu}. Rigol \emph{et al.} have indeed shown, using quantum Monte Carlo simulations, that confined Fermions interacting through a repulsive interaction may undergo an incompressible state characterized by full occupancy of the sites. When the confining harmonic potential is taken into account, the Mott-insulator phase is generally restricted to a finite domain at the center of the trap surrounded by compressible metallic phases \cite{rigol}. Liu \emph{et al.} have put forward the fact that collective oscillations of the atomic mass density constitute efficient signatures of the MMIT \cite{liu}. This result, based on the Bethe ansatz solution of the one-dimensional Fermi-Hubbard model, is currently motivating experimentalists to investigate the fermionic Mott-insulator phase in optical lattices \cite{ess1,ess2}. Very recently, the MMIT has been theoretically investigated for a fermionic system with interactions near the unitary limit \cite{bulgac,moon}. The competition between the BCS-superfluid and the insulator phases in strongly interacting Fermions systems has also been studied in Ref.\cite{zhai}. Finally the MMIT has been investigated in the context of three-dimensional optical traps \cite{rosch}. \\

In this context, it is interesting to study the behavior of the MMIT when the system rotates \cite{Holland1,Holland2}, or equally, when a light-induced effective ``magnetic" field is produced in the optical lattice \cite{Jaksch,Mueller,Demler,Ohberg,Ohberg2,porto}. Such artificial fields, which are currently realized in laboratories in order to simulate the dynamics of charged particles in a magnetic field, open the door for profound explorations of condensed matter physics \cite{porto}. In particular, optical lattices subjected to effective ``magnetic" fields constitute versatile systems which allows the investigation of vortex-physics \cite{goldman1,Holland3,wu,burkov,dan} and the quantum Hall effects \cite{goldman2,goldman3,Demler,Palmer,Hafezi}.  In the context of Bose gases, the superfluid-insulator phase transition is known to be affected by effective ``magnetic" fields \cite{oktel1,oktel2,dan}. Quite surprisingly, the phase boundary for this strongly-correlated system highly depends on the single-particle energy spectrum. The latter, originally studied in the context of electrons subjected to high magnetic fields by Hofstadter and Wannier, is fractal and its representation as a function of the effective ``magnetic flux" adopts the shape of an intriguing butterfly, the so-called Hofstadter butterfly \cite{hofs,wannier}. It has been proved recently that the phase boundary for the superfluid-insulator transition is directly related to the Hofstadter butterfly's edge \cite{oktel2,dan}. In a very different context, this highly irregular curve is also known to represent the phase boundary for the normal-superconducting phase transition in superconducting networks \cite{lopez,rammal,alex}. \\

In this work, we study the MMIT for a Fermi gas confined in a two-dimensional optical lattice. Contrary to the one-dimensional case, one has to treat the Fermi-Hubbard model on the basis of a mean-field approach. We first study the phase diagram in terms of the hopping amplitude and the chemical potential. We show that Mott-insulator phases, characterized by integer fillings and vanishing variance of the density, should be observed in two-dimensional fermionic optical lattices. In the presence of a harmonic trapping potential, we show that local Mott-insulator phases are generally present at the center of the trap, while the average variance is non-zero. We then study the system in the presence of an effective ``magnetic" field. In particular, we study the phase boundary in terms of the effective ``magnetic" flux and compare it to the single-particle spectrum.

\section{Mean field theory}

\begin{center} 
\begin{figure}
{\scalebox{0.47}{\includegraphics{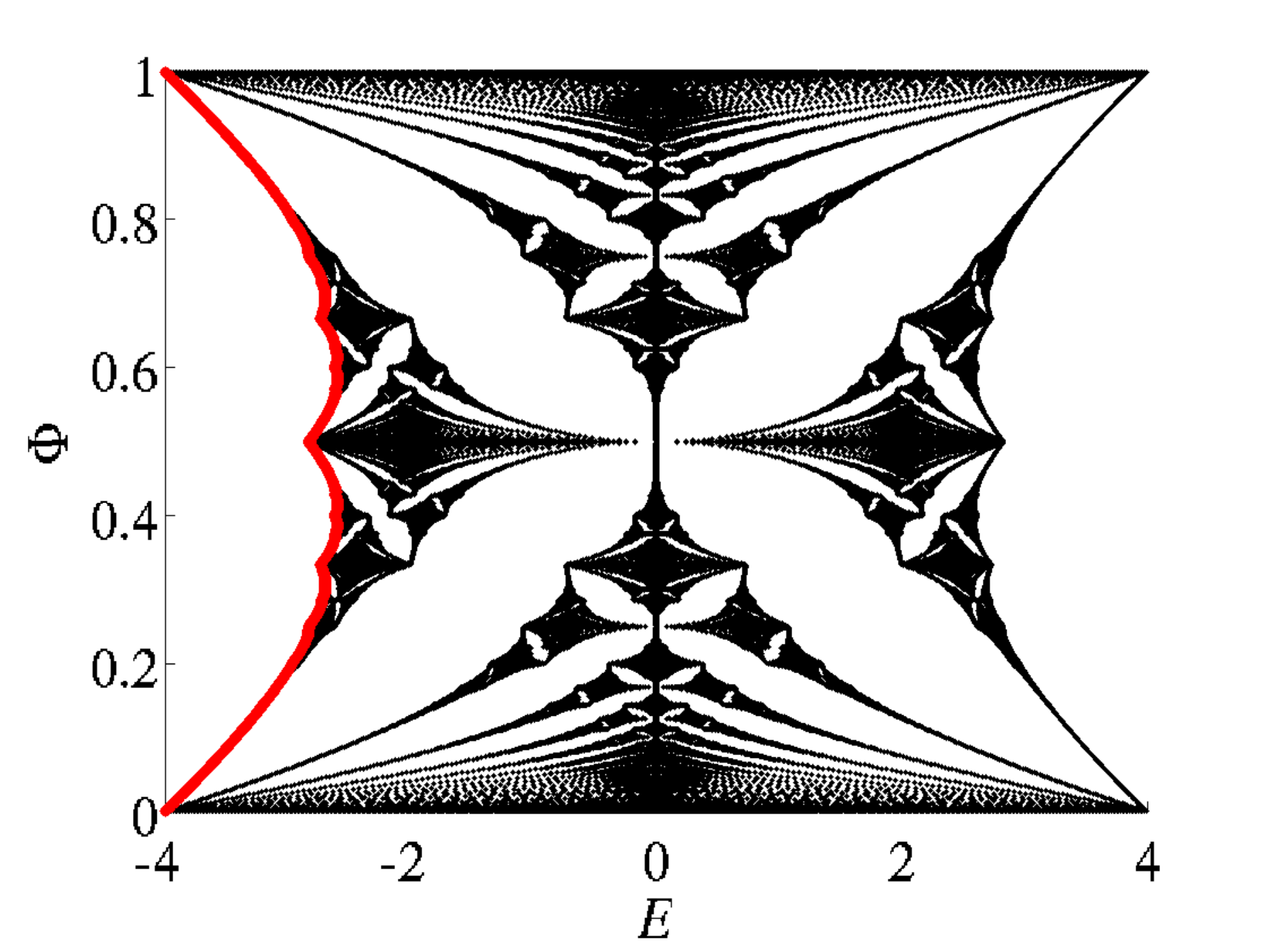}}} 
\caption{\label{square_but} (Color online) Hofstadter butterfly $\Phi= \Phi(E)$: single-particle spectrum for non-interacting Fermions in a rotating optical square lattice. The butterfly's edge is colored in red. }
\end{figure} 
\end{center}

We consider the case of a trapped fermionic gas in the presence of an effective ``magnetic" field. We assume that the optical potential created by the lasers is sufficiently strong in order to apply a tight-binding approximation. The effective ``magnetic" field $\boldsymbol{B}=B \, \boldsymbol{1_z}$, which is supposed to be created in the system with lasers \cite{Jaksch} or by rotation \cite{Holland2,Holland3}, is characterized by the parameter $\Phi$. The latter is the number of effective ``magnetic" flux quanta per unit cell. For a rotating system, this parameter is given by $\Phi=2 M a^2 \Omega /h$, where $\Omega$ is the angular velocity, $M$ is the particle's mass, $a$ is the lattice constant and $h=2 \pi \hbar$ is Planck's constant \cite{Holland2}. We can treat the many-body problem by considering the Landau gauge \cite{Jaksch}, $\boldsymbol{A}=(0, B x,0)$, in which the Fermi-Hubbard Hamiltonian reads 
\begin{align} 
\mathcal{H}=&  -t  \sum_{m,n , \sigma} f^{\dagger}_{m+1,n, \sigma} f_{m,n,\sigma}  + e^{i \theta (m)}f^{\dagger}_{m,n+1, \sigma} f_{m,n,\sigma} +h.c. \notag \\
&+U\sum_{m,n}  \hat{n}_{m,n,\uparrow} \hat{n}_{m,n,\downarrow}- \mu \sum_{m,n, \sigma}   \hat{n}_{m,n,\sigma} \notag \\
&+ \sum_{m,n, \sigma}  V_{\textrm{harm}}(m,n) \, \, \hat{n}_{m,n,\sigma} ,
\label{ham}
\end{align}
where $f_{m,n,\sigma}$ ($f^{\dagger}_{m,n, \sigma}$) is the fermionic annihilation (creation) operator of a particle on the lattice site $(m,n)$ with spin $\sigma (= \uparrow , \downarrow)$, and where the local density operator per spin is denoted by $\hat{n}_{m,n,\sigma}$. The fermionic operators satisfy the anticommutation rule $\{ f_{\xi},f^{\dagger}_{\xi'} \} _A=\delta (\xi-\xi')$. The parameter $t$ is the nearest-neighbor tunneling amplitude, $U$ is the repulsive interaction ($U>0$), $\mu$ is the chemical potential and $V_{\textrm{harm}}$ is a trapping potential \cite{ouradv}. In recent experiments involving trapped $^{40}$K, the optical lattice has lattice depth $V_0=5 E_R$, where $E_R=h^2/2M \lambda ^2$ is the recoil energy, $a=\lambda/2 =532$ nm is the lattice constant, and the tunneling amplitude $t \approx h \times 290$ Hz \cite{ess2}. The tight-binding regime is reached for $V_0/E_R > 2.23$ \cite{zhai}, which is the case for these experiments. In the following, we set the lattice constant $a$ and the particle's mass $M$ to unity and work in units where $\hbar=1$. \\
In the Landau gauge, the magnetic phase $\theta (m)$ is given by 
\be
\theta (m)= \int_{(m,n)}^{(m,n+1)} A_y (m) dy= 2 \pi \Phi m,
\ee
 where $\Phi=B/2\pi$ is the ``magnetic" flux quanta per unit cell and where the integral is performed along the link connecting the lattice sites $(m,n)$ and $(m,n+1)$  \cite{Jaksch}. We point out that a similar Fermi-Hubbard Hamiltonian can be obtained with the symmetric gauge $\boldsymbol{A}=\boldsymbol{x} \times \boldsymbol{\Omega}$ and can be found in Refs. \cite{Holland2,Holland3}.  In this work, the effective ``magnetic" flux $\Phi$ is supposed to vary between $[0,1]$. This ``high magnetic field regime" can easily be reached with light-induced gauge fields \cite{Jaksch}. In the context of rotating optical lattices, this regime is reached for $\Omega \approx 1$ kHz.  \\
In the following we consider that the ``magnetic" flux is rational, namely $\Phi=p/q$ where $p$ and $q$ are integers. In this case, the single-particle Schr\"odinger equation (for $U=V_{\textrm{harm}}=0$) yields the well-known Harper equation \cite{hofs}. This difference equation is expressed as 
\begin{equation}
e ^{i k _x} \, \psi_{m+1} + e ^{-i k _x} \, \psi_{m-1}  + 2 \cos (2 \pi \Phi m + k _y) \, \psi_m = E  \, \psi_m ,
\label{harp}
\end{equation}
where $\psi_{m}$ is a $q$-periodic wave function, $\boldsymbol{k}$ is the wave vector and $E$ is the single-particle energy. The wave vector belongs to the magnetic Brillouin zone, a two-torus defined as $k_x \in [0, \frac{2 \pi}{q}]$ and  $k_y \in [0,2 \pi]$. The energy spectrum associated to Eq. \eqref{harp} has a band structure, consisting of $q$ subbands, which has been extensively studied in the literature \cite{hofs,wannier,rudinger}. The representation of the energy as a function of the ``magnetic" flux $\Phi$ leads to a well known fractal, the so-called Hofstadter butterfly (Fig.\ref{square_but}). \\
In order to study the interacting system Eq. \eqref{ham}, we approximate the hopping term on the basis of the decoupling formula for the product of two field operators \cite{sheshadri,oktel1,goldman1},
\begin{align}
f_{m,n,\sigma} f_{m+1,n,\sigma}=&\langle f_{m,n,\sigma} \rangle f_{m+1,n,\sigma}+f_{m,n,\sigma} \langle f_{m+1,n,\sigma} \rangle  \notag \\
& - \langle f_{m,n,\sigma} \rangle \langle f_{m+1,n,\sigma} \rangle .
\label{approx}
\end{align}
This approximation holds for small fluctuations defined as $\delta_{m,n,\sigma}=\langle f_{m,n,\sigma}^2 \rangle-\langle f_{m,n,\sigma}\rangle ^2$. The mean-field Hamiltonian $\mathcal{H}_{\textrm{MF}}$ then takes the form
\begin{align} 
\mathcal{H}_{\textrm{MF}}=&  -t  \sum_{m,n , \sigma} \biggl ( f^{\dagger}_{m+1,n, \sigma} \Psi^*_{m,n,\sigma} +\Psi_{m+1,n, \sigma} f_{m,n,\sigma} \notag \\
&+ e^{i \theta (m)} \bigl ( \Psi_{m,n+1, \sigma} f_{m,n,\sigma} + f^{\dagger}_{m,n+1, \sigma} \Psi^*_{m,n,\sigma} \bigr ) +h.c.  \biggr ) \notag \\
&+U\sum_{m,n}  \hat{n}_{m,n,\uparrow} \hat{n}_{m,n,\downarrow}- \mu \sum_{m,n, \sigma}   \hat{n}_{m,n,\sigma} \notag \\
&+ \sum_{m,n, \sigma}  V_{\textrm{harm}}(m,n) \, \, \hat{n}_{m,n,\sigma} +t \, C_{m,n,\sigma},
\label{ham2}
\end{align}
where 
\begin{align} 
C_{m,n,\sigma}=& \Psi_{m+1,n,\sigma} \Psi^*_{m,n,\sigma} +e^{i \theta (m)} \Psi_{m,n+1,\sigma} \Psi^*_{m,n,\sigma}  \notag \\
&+\Psi_{m-1,n,\sigma} \Psi^*_{m,n,\sigma} +e^{-i \theta (m)} \Psi_{m,n-1,\sigma} \Psi^*_{m,n,\sigma} ,
\end{align}
and $\Psi_{m,n,\sigma}= \langle f^{\dagger}_{m,n,\sigma} \rangle$. \\
When $V_{\textrm{harm}}(m,n)=0$, the Hamiltonian is invariant with respect to discrete translations along the $y$-axis and the system adopts a strip geometry, namely $f_{m,n,\sigma}=f_{m,\sigma}$. Since the flux $\Phi=p/q$ is rational, one may set $f_{m,\sigma}=f_{m+q,\sigma}$ and the 2D system reduces to a $q \times 1$ supercell. In order to study the effects induced by a harmonic trap, we consider that the harmonic potential has its minimum at the centre of the supercell and is expressed as
\be
V_{\textrm{harm}}(m)=\biggl ( \frac{2}{q} \biggr )^2 V_2 \biggl ( m-\frac{q}{2} \biggr )^2 ,
\ee
inside the supercell, where $V_2$ is the potential strength. \\
The mean-field Hamiltonian is given by a sum of single-site terms $\mathcal{H}_{\textrm{MF}}=\sum_m \mathcal{H}_{\textrm{MF}}^m$ which are represented by $4 \times 4$ matrices in the occupation number basis $\vert N_{m, \uparrow} N_{m, \downarrow} \rangle$, where $N_{m, \sigma}=0,1$. The non-zero matrix elements are
\begin{align}
&\langle 0, 0 \vert \mathcal{H}_{\textrm{MF}}^m \vert 0, 0 \rangle =C_m , \notag \\
&\langle 1, 0 \vert \mathcal{H}_{\textrm{MF}}^m \vert 1, 0 \rangle =\langle 0 ,1 \vert \mathcal{H}_{\textrm{MF}}^m \vert 0, 1 \rangle =C_m - \mu +V_{\textrm{harm}}(m), \notag \\
&\langle 1 ,1 \vert \mathcal{H}_{\textrm{MF}}^m \vert 1, 1 \rangle =C_m - 2 \mu +2 V_{\textrm{harm}}(m) +U, \notag \\
& \langle 0 ,0 \vert \mathcal{H}_{\textrm{MF}}^m \vert 1, 0 \rangle = -t \bigl ( \Psi_{m+1, \uparrow}+ \Psi_{m-1, \uparrow}+2 \, \textrm{Cos} \, \theta(m)  \Psi_{m, \uparrow}  \bigr ) , \notag \\
& \langle 0 ,0 \vert \mathcal{H}_{\textrm{MF}}^m \vert 0, 1 \rangle = -t \bigl ( \Psi_{m+1, \downarrow}+ \Psi_{m-1, \downarrow}+2 \,  \textrm{Cos} \, \theta(m)  \Psi_{m, \downarrow} \bigr ) , \notag \\
& \langle 0 ,0 \vert \mathcal{H}_{\textrm{MF}}^m \vert 1, 0 \rangle =  \langle 0 ,1 \vert \mathcal{H}_{\textrm{MF}}^m \vert 1, 1 \rangle , \notag \\
&\langle 0 ,0 \vert \mathcal{H}_{\textrm{MF}}^m \vert 0, 1 \rangle =  \langle 1 ,0 \vert \mathcal{H}_{\textrm{MF}}^m \vert 1, 1 \rangle ,
\label{themat}
\end{align}
where 
\begin{align}
C_m=& \, t \sum_{\sigma}\biggl ( \Psi_{m+1,\sigma} \Psi^*_{m,\sigma} + \Psi_{m-1,\sigma} \Psi^*_{m,\sigma} \notag \\
&+ 2 \,  \textrm{Cos} \, \theta(m) \vert \Psi_{m,\sigma} \vert ^2 \biggr ),
\end{align}
and $\theta(m)= 2 \pi \Phi m$. \\
The physical quantities of interest for this work are the local density $\rho_m = \langle G_m \vert \hat{n}_{m} \vert G_m \rangle$, its variance $\Delta_m = \langle \hat{n}_{m} ^2 \rangle - \langle \hat{n}_{m}  \rangle ^2 $ and the average density $\rho=(1/q) \sum_m \rho_m$, where we have introduced the notation $\hat{n}_{m}= \sum_{\sigma} \hat{n}_{m,\sigma}$. We note that $\rho_m \le 2$ since there are two spin states. The ground state $\vert G_m \rangle$ is written as
\be
\vert G_m \rangle = \sum_{N_{\uparrow},N_{\downarrow}=0,1} \alpha^{N_{\uparrow},N_{\downarrow}}_m \vert N_{m, \uparrow} N_{m, \downarrow} \rangle ,
\label{ground}
\ee
where $\alpha_m=(\alpha_m^{(0,0)},\alpha_m^{(1,0)},\alpha_m^{(0,1)},\alpha_m^{(1,1)})$ is the eigenvector corresponding to the lowest eigenvalue of the matrix Eq.\eqref{themat}. The average values $\Psi_{m,\sigma}= \langle G_m \vert f^{\dagger}_{m,\sigma} \vert G_m \rangle$ are given by
\begin{align}
&\Psi_{m,\uparrow}=\bigl ( \alpha_m^{(1,0)} \bigr )^*  \alpha_m^{(0,0)} +\bigl ( \alpha_m^{(1,1)} \bigr )^*  \alpha_m^{(0,1)} , \notag \\
&\Psi_{m,\downarrow}=\bigl ( \alpha_m^{(0,1)} \bigr )^*  \alpha_m^{(0,0)} +\bigl ( \alpha_m^{(1,1)} \bigr )^*  \alpha_m^{(1,0)}.
\label{thepsis}
\end{align}
Eqs. \eqref{themat} , \eqref{ground} and \eqref{thepsis} form a set of self-consistent equations, which are solved in order to evaluate the ground state, the average values $\Psi_{m,\sigma}$ and the physical quantities $\rho$, $\rho_m$ and $\Delta_m$.\\
The local density $\rho_m$ is expressed as
\begin{align}
 \rho_m&= \langle G_m \vert \hat{n}_{m} \vert G_m \rangle \notag \\
&= \vert \alpha_m^{(0,1)} \vert ^2 + \vert \alpha_m^{(0,1)} \vert ^2 + 2 \vert \alpha_m^{(1,1)} \vert ^2 ,
\end{align}
the local density's variance $ \Delta_m$ is given by
\begin{align}
 \Delta_m &= \langle G_m \vert \hat{n}_{m} ^2 \vert G_m \rangle -  \rho_m^2 \notag \\
& =  \vert \alpha_m^{(0,1)} \vert ^2 + \vert \alpha_m^{(0,1)} \vert ^2 + 4 \vert \alpha_m^{(1,1)} \vert ^2   \notag \\
& -  \biggl ( \vert \alpha_m^{(0,1)} \vert ^2 + \vert \alpha_m^{(0,1)} \vert ^2 + 2 \vert \alpha_m^{(1,1)} \vert ^2 \biggr )^2 
\end{align}
and the average density $\rho$ is evaluated according to
\begin{align}
 \rho&=\frac{1}{q} \sum_m \langle G_m \vert \hat{n}_{m} \vert G_m \rangle \notag \\
&= \frac{1}{q} \sum_m \vert \alpha_m^{(0,1)} \vert ^2 + \vert \alpha_m^{(0,1)} \vert ^2 + 2 \vert \alpha_m^{(1,1)} \vert ^2 .
\end{align}

\section{Numerical investigation}

\subsection{Phase diagram in the $\mu - t$ plane for the homogeneous system}

\begin{center} 
\begin{figure}
{\scalebox{0.38}{\includegraphics{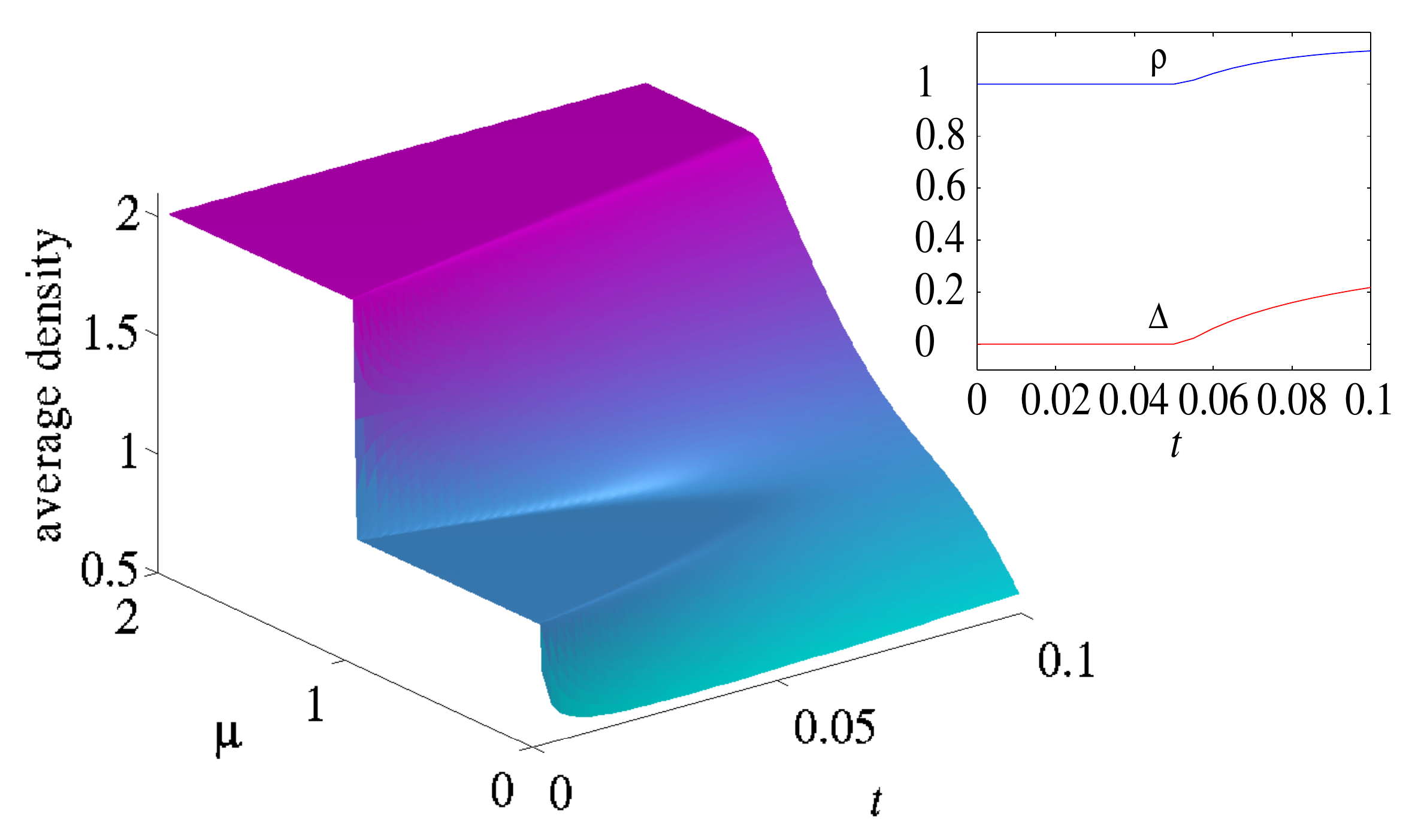}}} 
\caption{\label{two_lobes} (Color online) Average density $\rho =\rho \, (t, \mu)$ in the $t-\mu$ plane, for $V_2=\Phi=0$. Mott-insulator and band-insulator phases, corresponding to integer fillings, depict two distinctive plateaus at $\rho=1$ and $\rho=2$. In the right corner: average density  $\rho$ (blue) and average variance  $\Delta$ (red) as a function of the hopping amplitude $t$, for $\mu=0.7$. Inside the  Mott lobe $\rho=1$ and $\Delta =0$. The hopping parameter $t$ and the chemical potential $\mu$ are expressed in units of the on-site interaction energy $U$.}
\end{figure} 
\end{center}

In the absence of the harmonic trap $V_2=0$, the system is invariant under discrete translations along the $x$ direction. As a result, the local density $\rho_m=\rho$ and its variance $\Delta_m=\Delta$ are constant in the supercell. In this case, the representation of the average density $\rho$ in the $t-\mu$ plane shows two distinctive plateaus (see Fig.\ref{two_lobes}). The latter correspond to the Mott-insulator (resp. band-insulator) phases which are characterized by integer fillings, $\rho=1$ (resp. $\rho=2$), and by a vanishing variance $\Delta=0$. Contrary to the predictions obtained on the basis of quantum Monte-Carlo simulations  for the 1D fermionic optical lattice \cite{rigol}, the variance $\Delta$ evaluated within our mean-field theory  plays the role of an order parameter for the metal-insulator transition. In the right corner of Fig.\ref{two_lobes}, we represent the density $\rho$ and the variance $\Delta$ as a function of the hopping parameter $t$: inside the Mott lobe ($t < 0.05$ for $\mu=0.7$), $\rho=1$ and the variance indeed vanishes $\Delta =0$. Outside of this Mott-insulator domain, the variance has a finite value which increases as the hopping amplitude intensifies and the average density $\rho>1$.\\
In order to justify the mean-field approximation Eq. \eqref{approx}, we have evaluated the fluctuations $\vert \delta_{m,\sigma} \vert= \vert \langle f_{m,\sigma}^2 \rangle-\langle f_{m,\sigma}\rangle ^2 \vert $, and we have verified that $\vert \delta_{m,\sigma} \vert \ll \langle f_{m,\sigma}\rangle $ for an important region surrounding the Mott lobes. It is worth noticing that $\langle f_{m,\sigma}\rangle$ is not an order parameter for the MMIT since it takes a non-zero value in a small region within the borders of the Mott lobes. \\

\subsection{In the presence of a harmonic trap}
\begin{center} 
\begin{figure}
{\scalebox{0.37}{\includegraphics{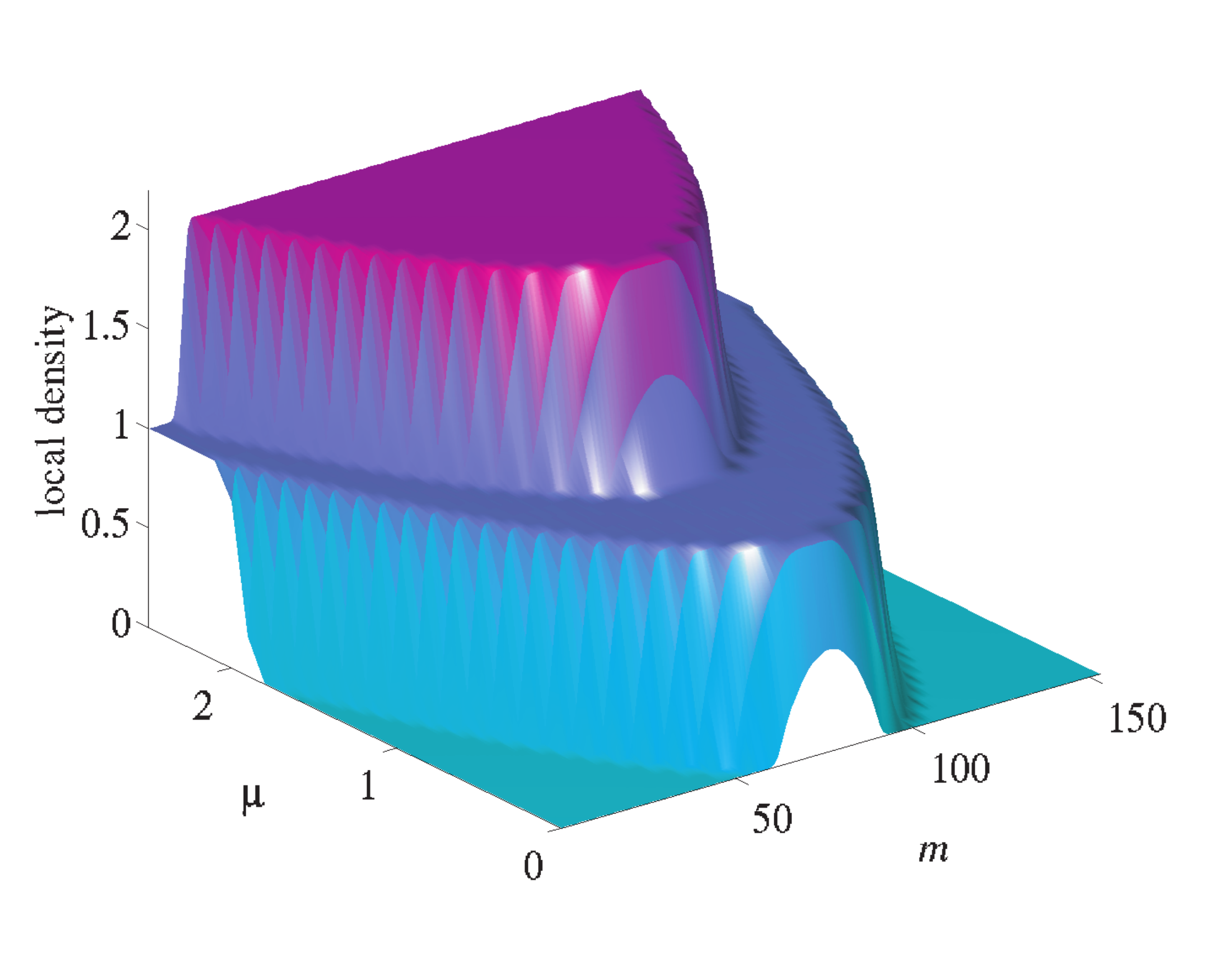}}} 
\caption{\label{mudens} (Color online) Local density $\rho_m=\rho_m(\mu)$ as a function of the chemical potential $\mu$, for  $t=0.02 \, U$, $V_2=2 \, U$ and for $q=153$ lattice sites in the supercell. The  chemical potential $\mu$ is expressed in units of the on-site interaction energy $U$. }
\end{figure} 
\end{center}

\begin{center} 
\begin{figure}
{\scalebox{0.4}{\includegraphics{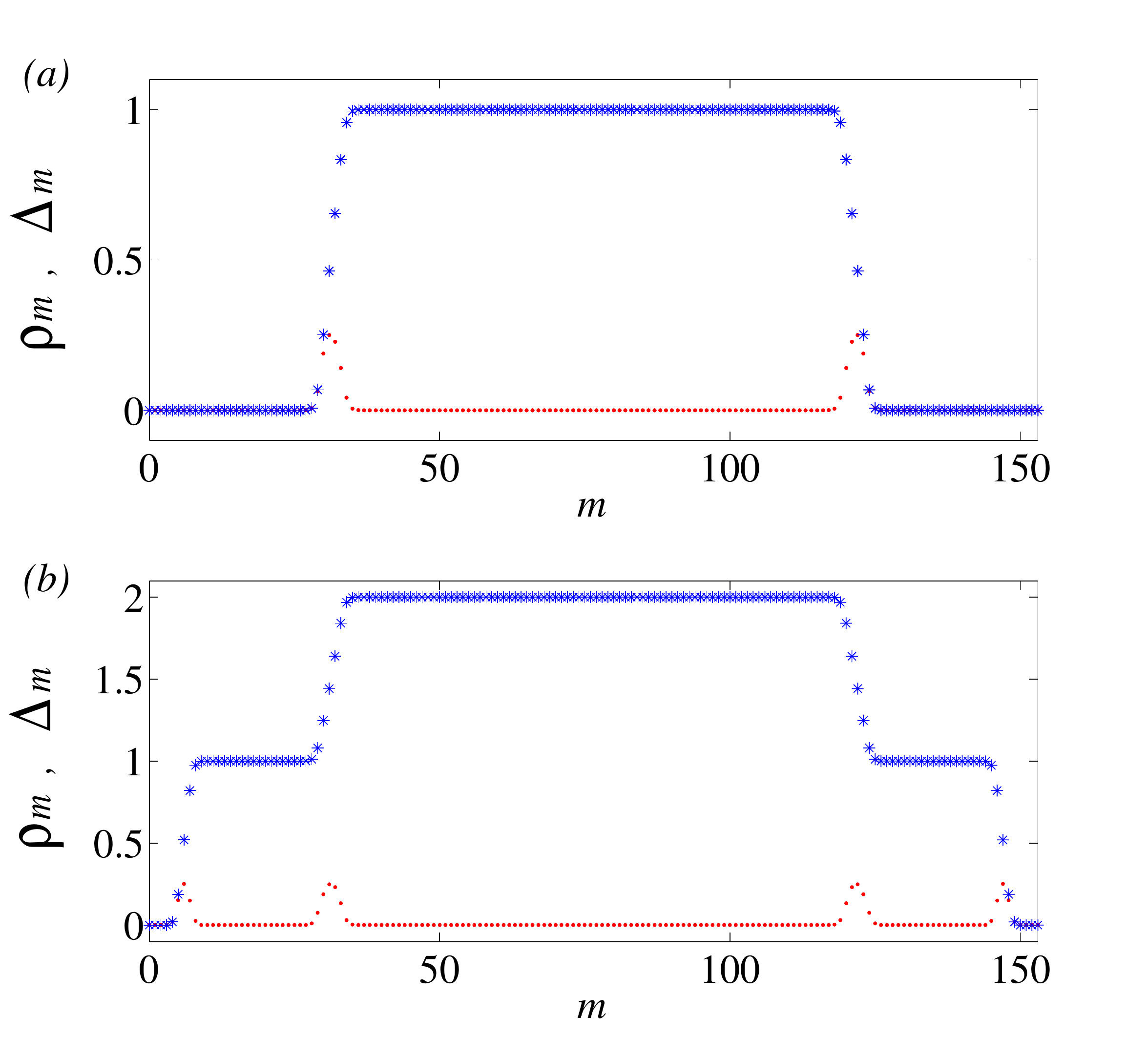}}} 
\caption{\label{dens} (Color online) Local density $\rho_m$ (blue asterisk) and the local variance $\Delta_m$ (red dot) when $V_2=2 \, U$:  $(a)$ $\mu=0.7 \, U$ and $(b)$ $\mu=1.7 \, U$. We set $t=0.02 \, U$ and $q=153$ in this figure.}
\end{figure} 
\end{center}

\begin{center} 
\begin{figure}
{\scalebox{0.3}{\includegraphics{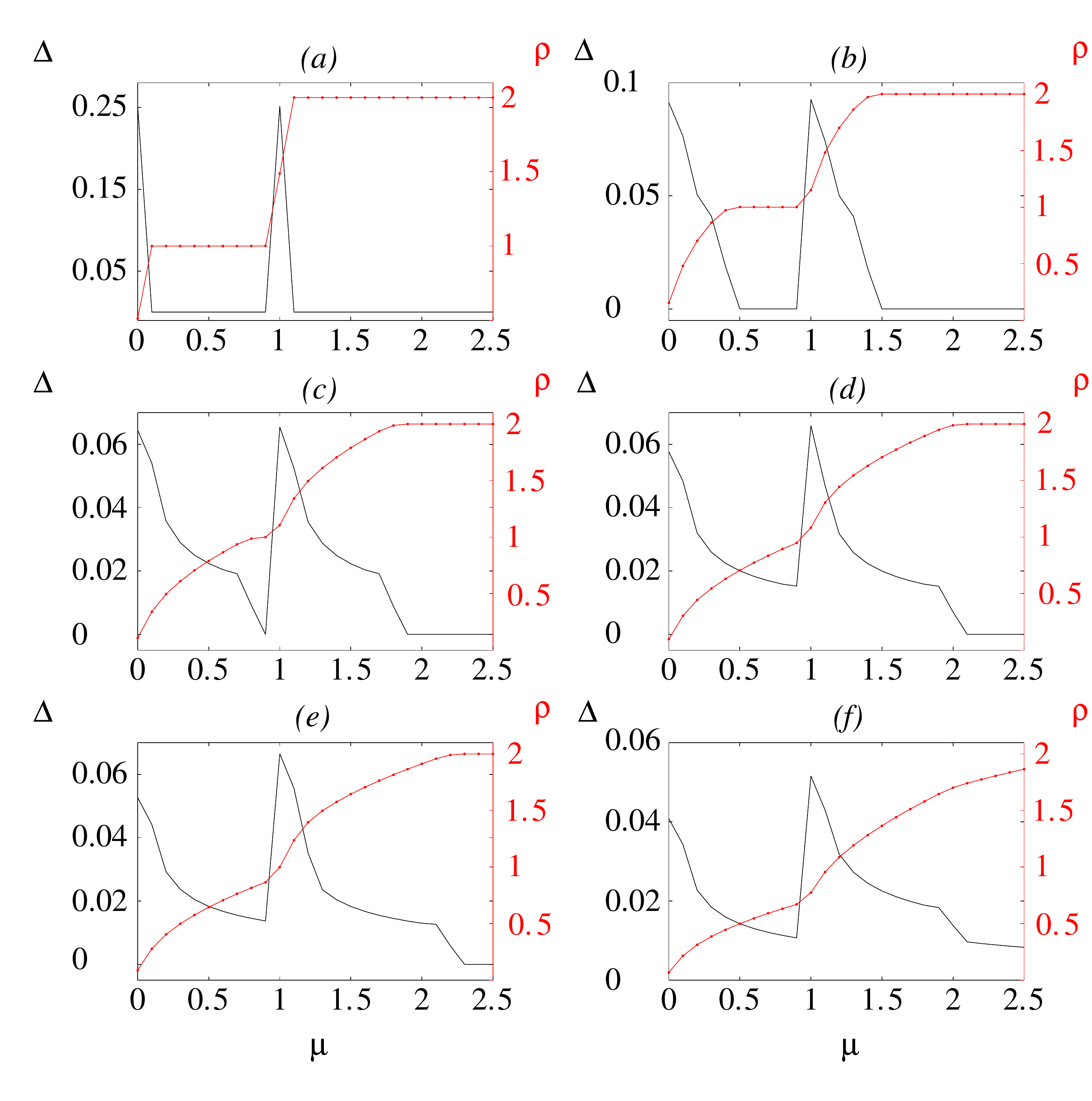}}} 
\caption{\label{av_dens} (Color online) Average density $\rho= \frac{1}{q} \sum_m \rho_m$ (red dotted line) and average variance $\Delta = \frac{1}{q} \sum_m \Delta_m$ (black line) as a function of the chemical potential $\mu$, for $t=0.02 \, U$ and $q=153$: $(a)$ $V_2=0$, $(b)$ $V_2=0.4 \, U$, $(c)$ $V_2=0.8 \, U$, $(d)$ $V_2=1 \, U$, $(e)$ $V_2=1.2 \, U$ and $(f)$ $V_2=2 \, U$. Global Mott-insulator (resp. band-insulator) phases correspond to integer fillings $\rho=1$ (resp. $\rho=2$) and vanishing variance $\Delta=0$. The chemical potential $\mu$ is expressed in units of the on-site interaction energy $U$.  }
\end{figure} 
\end{center}

When the harmonic trap is added within the supercell $m \in [1,q]$, the particles agglomerate at the center of the trap and the local density is not necessarily constant. In order to study this effect, we show the local density $\rho_m$ as a function of the chemical potential $\mu$ in Fig.\ref{mudens}. We note that for sufficiently large supercells ($q >50$), the results obtained in this section aren't affected by the specific value given to the parameter $q$. Therefore we choose an arbitrary value and we set $q=153$, the computations are thus performed in a $153 \times 1$ supercell, and we fix the hopping parameter $t=0.02$. In accordance with the results obtained with quantum Monte-Carlo simulations for the 1D fermionic optical lattice \cite{rigol}, we observe the emergence of local Mott-insulator phases in the center of the system as the chemical potential increases. In Fig.\ref{mudens}, which corresponds to the case $V_2=2$, one observes two distinctive plateaus of Mott-insulator and band-insulator phases. For small values of the chemical potential ($\mu<0.098$ for $V_2=2$), we observe a parabolic density profile which points out that the whole system is in a metallic phase \cite{rigol,liu}. When the chemical potential varies between $0.098 < \mu < 0.91$, the system is characterized by a  Mott-insulator phase with $\rho_m=1$, localized at the center of the trap and surrounded by two metallic phases. For $0.91 < \mu < 1.08$, a localized metallic phase develops in the middle of the Mott-insulator phase. When the chemical potential is greater than $\mu > 1.08$, a band-insulator phase with $\rho_m=2$ appears at the center of the system: the latter is then characterized by three distinctive insulator phases, which are generally separated and surrounded by metallic phases. We point out that such a behaviour also occurs in the one-dimensional case: Fig.\ref{mudens} reproduces the density profiles obtained with quantum Monte Carlo simulations (see Fig. 1 and Fig. 2 of Ref. \cite{rigol}), and with the Bethe ansatz (see Fig. 1 of Ref. \cite{liu}).
In order to identify the local insulator phases illustrated in Fig.\ref{mudens}, we represent the local density $\rho_m$ and the variance $\Delta_m$ for two fixed values of the chemical potential ($\mu=0.7$ and $\mu=1.7$) in Fig.\ref{dens}. In this figure, one observes that a Mott plateau indeed develops at the center of the trap for $\mu=0.7$ (see Fig.\ref{dens} $(a)$). This plateau is characterized by $\rho_m=1$ and by a vanishing variance $\Delta_m=0$. We notice that the variance $\Delta_m$ is non-zero in the metallic region, where $\rho_m$ is not an integer. When $\mu=1.7$, the central plateau corresponds to $\rho_m=2$, while two plateaus at $\rho_m=1$ remain near the edges (see Fig.\ref{dens} $(b)$). The metallic regions, which surround each insulator plateau is characterized by a non-vanishing variance $\Delta_m$. \\
We also notice that the density vanishes at the edges of the supercell for $\mu<1.8$, according to the fact that the particles concentrate at the center of the harmonic trap. In fact, as the chemical potential increases, the edges undergo different phases: for $1.8<\mu<2.1$ the edges are in a metallic phase, for $2.1<\mu<2.8$ the edges are in a Mott-insulator phase with $\rho_m=1$, for $2.8<\mu<3.1$ the edges are again in a metallic phase, and for $\mu>3.1$ the whole system is in a global band-insulator phase with $\rho_m=2$.
As the strength of the harmonic potential $V_2$ increases, the global phase diagram $\rho=\rho (t , \mu)$, represented in Fig.\ref{two_lobes} for $V_2=0$, varies: the first Mott lobe ($\rho=1$) progressively diminishes and eventually disappears for $V_2 >1$, while the second insulator region ($\rho=2$) appears at increasing values of the chemical potential. This effect is illustrated in Fig.\ref{av_dens}, which shows the average density $\rho = \frac{1}{q} \sum_m \rho_m$ and the average variance $\Delta = \frac{1}{q} \sum_m \Delta_m$ as a function of the chemical potential, for various values of the potential strength $V_2 \in [0, 2]$. Fig.\ref{av_dens} is actually a cut through the $\mu-t$ phase diagram, at  a fixed hopping parameter $t=0.02$. In this figure, one observes that for $V_2 < 1$, the system reaches the two insulator domain as the chemical potential increases. The Mott-insulator and band-insulator phases are characterized by a vanishing average variance $\Delta$. For sufficiently small potential strength, the function $\rho=\rho (\mu)$ has two plateaus at $\rho=1$ and $\rho = 2$. The first plateau progressively diminishes while $V_2$ is increased (see Fig.\ref{av_dens} $a$, Fig.\ref{av_dens} $b$ and Fig.\ref{av_dens} $c$, corresponding to $V_2=0$, $V_2=0.4$ and $V_2=0.8$ respectively). When $V_2 \ge1$, the function $\rho=\rho (\mu)$ only reaches one plateau at $\rho =2$, for $\mu >2$ (see Fig.\ref{av_dens} $d$, Fig.\ref{av_dens} $e$ and Fig.\ref{av_dens} $f$, corresponding to $V_2=1$, $V_2=1.2$ and $V_2=2$ respectively). \\
It is worth noticing that a system with non-vanishing average variance $\Delta$ may exhibit local insulator phases inside the trap. This is the case for the specific situation illustrated in Fig.\ref{dens} $(b)$ which exhibits two plateaus at $\rho_m=1$ and $\rho_m=2$, while $\rho \approx 1.5$ and $\Delta \approx 0.02$ (for $\mu=1.7$). We finally note the interesting fact that if the average variance is vanishing, $\Delta=0$, then the local density $\rho_m$ is constant ($\rho_m$ gives the integer filling of the whole lattice) in spite of the inhomogeneous confining potential. We can conclude that a global Mott-insulator phase enhances a homogenization of the system.

\subsection{Mott metal-insulator phase transition in the presence of  a ``magnetic" field}
\begin{center} 
\begin{figure}
{\scalebox{0.27}{\includegraphics{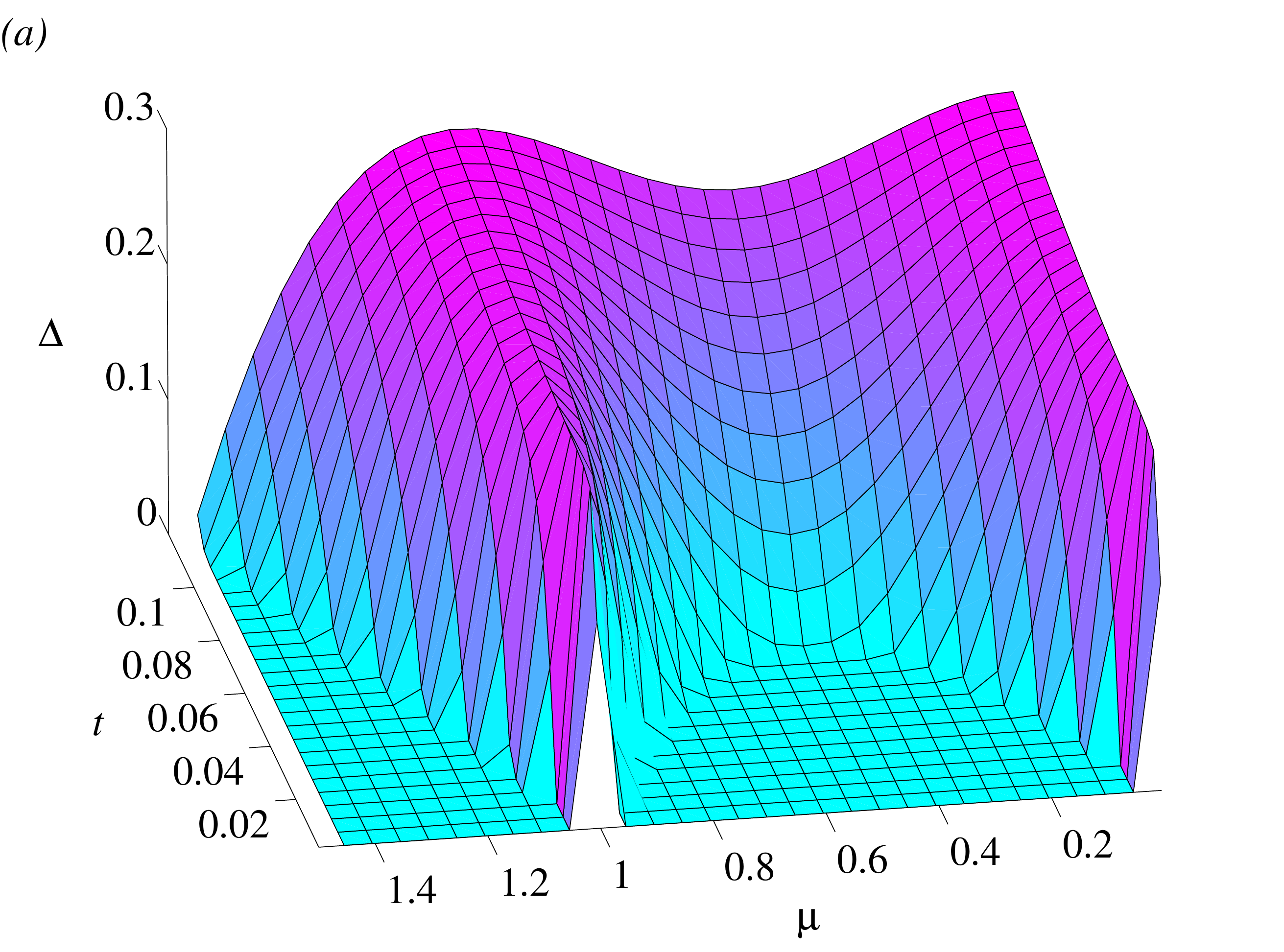}}} 
{\scalebox{0.27}{\includegraphics{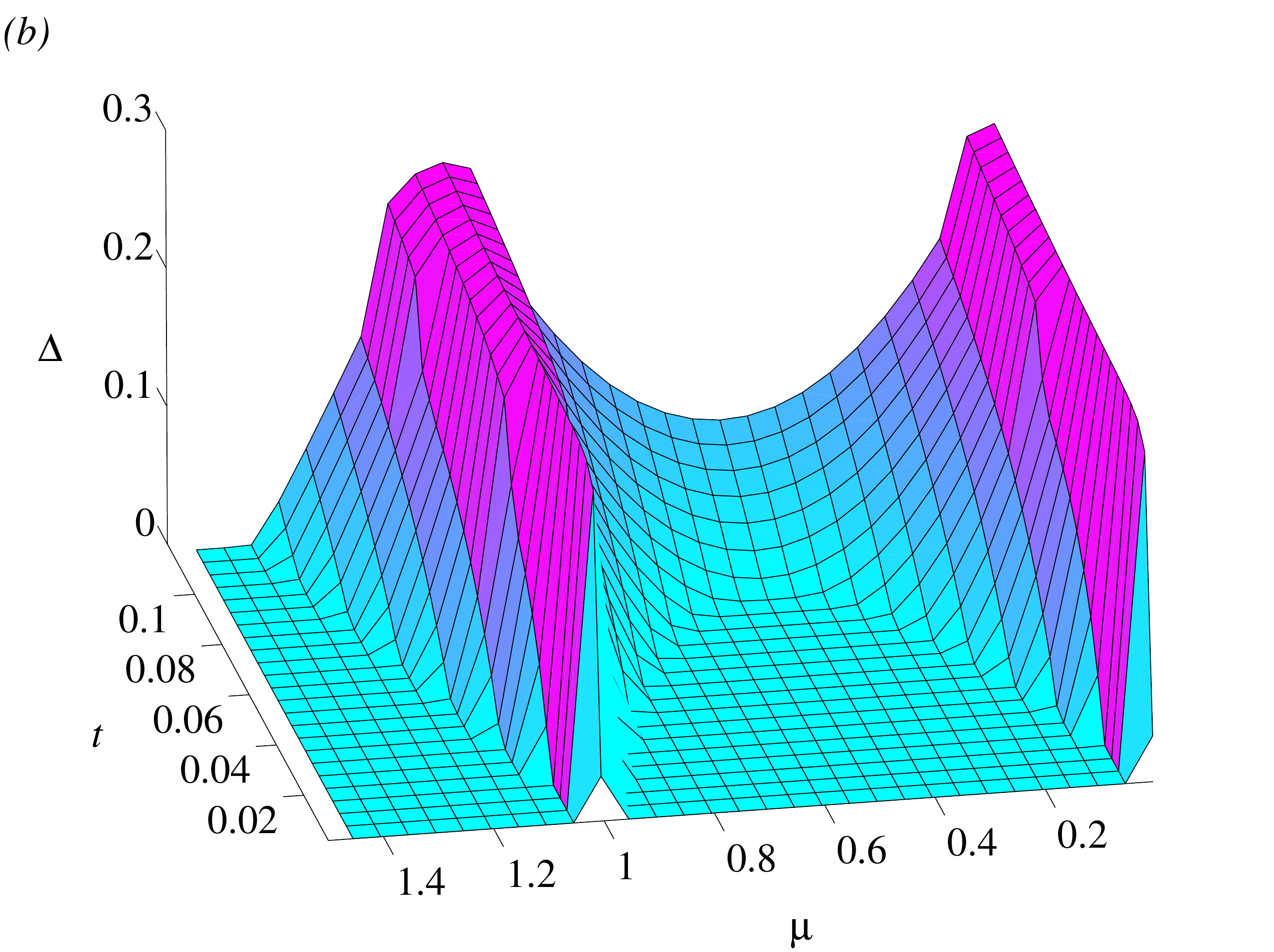}}} 
\caption{\label{comp_phi} (Color online) Average variance $\Delta=\frac{1}{q} \sum_m \Delta_m$ as a function of the hopping parameter $t$ and the chemical potential $\mu$, for $V_2=0$: $(a)$ $\Phi=0$ and $(b)$ $\Phi=0.2$. The hopping parameter $t$ and the chemical potential $\mu$ are expressed in units of the on-site interaction energy $U$. }
\end{figure} 
\end{center}

\begin{center} 
\begin{figure}
{\scalebox{0.43}{\includegraphics{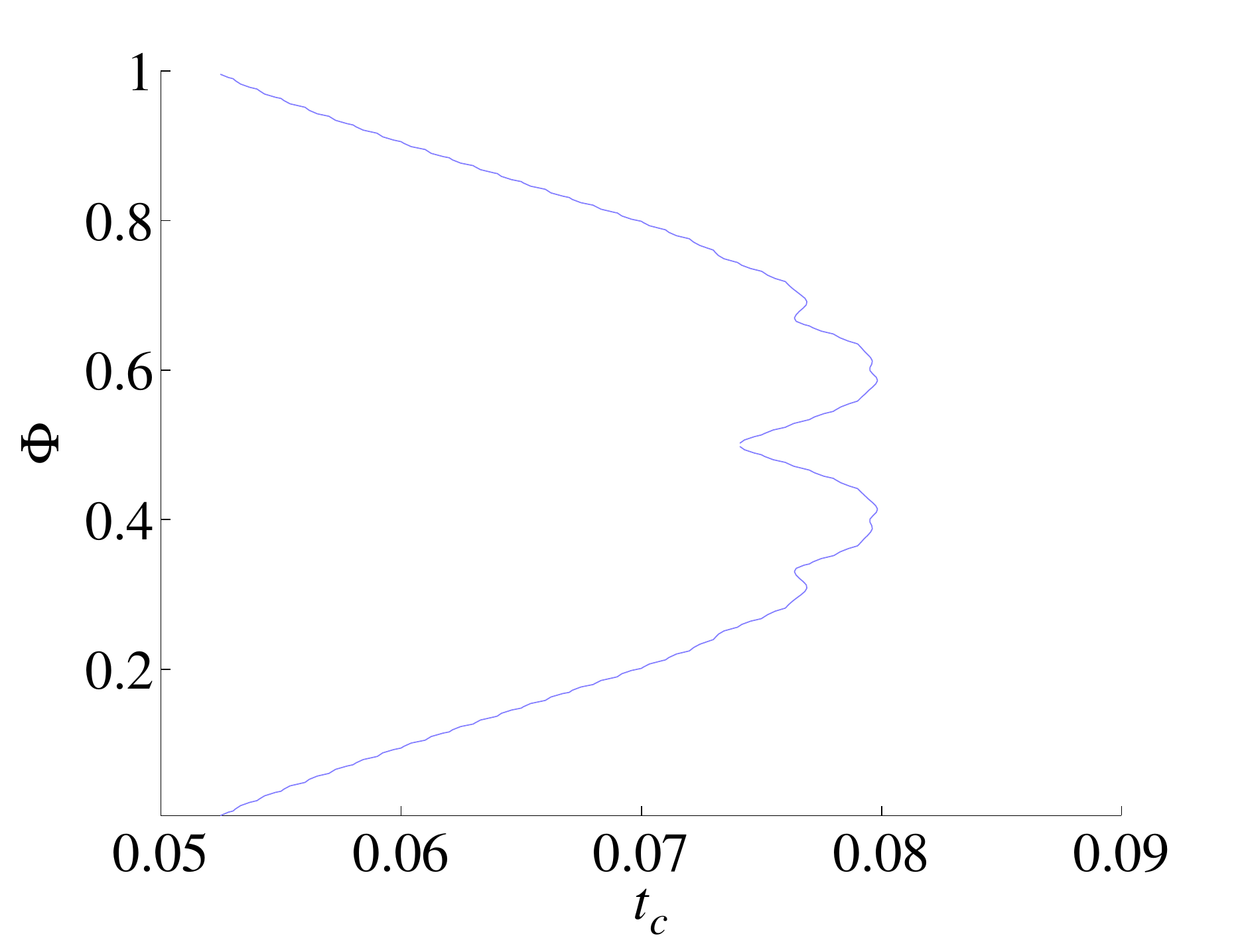}}} 
\caption{\label{hof_bord} (Color online) Critical hopping parameter $t_c$ as a function of the effective ``magnetic" flux $\Phi=p/q$, for $q=277$ and the chemical potential $\mu=0.3 \, U$. In this figure, we set $V_2=0$. The hopping parameter $t$ is expressed in units of the on-site interaction energy $U$.}
\end{figure} 
\end{center}

The mean-field theory developed in the previous section allows one to investigate the metal-insulator phase transition in a wide range of the parameters. Besides, one of the great advantages of our theory resides in the fact that we can investigate the MMIT when the two-dimensional system rotates. In particular, the ``high magnetic field" regime $\Phi \approx 1$, can be considered. \\
The MMIT phase diagram, represented in the $\mu - t$ plane, is affected by the presence of a high ``magnetic" field. In particular, the Mott lobe corresponding to $\rho=1$ extends over greater values of the hopping parameter $t$. We show this phenomenon in Fig.\ref{comp_phi}, which illustrates the phase diagram for $\Phi=0$ and $\Phi=0.2$ in terms of the average variance $\Delta$ (the Mott lobes are designated by $\Delta=0$). For $\Phi=0$, the Mott-lobe extends to the critical value $t_c \approx 0.06$, whereas it reaches the critical value $t_c \approx 0.08$ for $\Phi=0.2$. We also note that the band-insulator domain ($\rho=2$) is modified by the ``magnetic" field: for $\mu=1.3$, one observes a critical value $t_c \approx 0.07$ for $\Phi=0$, whereas one finds $t_c \approx 0.1$ for $\Phi=0.2$. Finally we point out that the insulator lobes are only modified according to the value of the ``magnetic" flux $\Phi =p/q$, and that they are not affected by the supercell length $q$. 
A richer representation of this effect is drawn in Fig.\ref{hof_bord} which shows the critical hopping parameter $t_c$ as a function of the flux $\Phi=p/q$ for a fixed value of the chemical potential (we set $\mu=0.3$). Here we set $q=227$, so that $p$ and $q$ are mutally primes for all $p=1, 2,... q$ \cite{rmk}. The phase boundary $t_c=t_c(\Phi)$ is highly irregular and exhibits pronounced slopes at $\Phi=1/2$, $\Phi=1/3$ and $\Phi=2/5$. This figure is symmetric with respect to $\Phi=0.5$ and is periodic with period $T_{\Phi}=1$. The general shape of the phase boundary $t_c=t_c(\Phi)$ is similar for all the values of the chemical potential: in particular, the MMIT which takes place around the band-insulator domain ($\rho=2$) is also described by such a curve. \\
In fact, this important numerical result reveals that the phase boundary for the MMIT is directly related to the edge of the Hofstadter butterfly represented in Fig.\ref{square_but}. The fact that this phase transition is intimately connected to the underlying single-particle physics is not too surprising: in the context of rotating Bose gases, it is already known that the phase boundary $t_c=t_c(\Phi)$ for the superfluid-Mott insulator transition also displays the Hofstadter butterfly's edge \cite{dan,oktel2}. We emphasize here that our numerical results suggest that this fact holds for rotating fermions. This important result enforces the idea that rotations, or artificial ``magnetic" fields, induces a modification to the MMIT in a way which is independent of the particles statistics. Moreover the universal character of this modification is dictated by the Hofstadter butterfly. As already discussed in Ref. \cite{dan}, it should be noted that most of the atoms are static in the vicinity of the insulator domains.  The gas of mobile particles and holes is thus highly diluted in these regions. This remark suggests that the system behaves according to single-particle physics around the insulator regions and therefore partially explains why these phase boundaries may be related to the fractal single-particle spectrum.  \\
Finally, we note that when the system reaches the metallic phase, $\Delta \ne 0$,  local insulator phases may still be present in the system because of the presence of a ``magnetic" field. Contrary to the effects induced by a trapping potential and described in the previous section, these local Mott-insulator phases are generally periodically distributed in space.

\section{Conclusion}

In this work, we have studied an important feature of ultracold Fermions physics. We have shown that two-dimensional fermionic optical lattices should exhibit a Mott metal-insulator phase transition. In the presence of a trapping potential, we have observed the formation of local Mott-insulator phases, generally situated at the center of the trap. These local Mott-insulator phases are characterized by a vanishing variance of the local density. When the system is subjected to an external ``magnetic" field, or equally when the system rotates, we have put forward the fact that the Mott lobes are sensible to the specific value of the effective ``magnetic" flux: for a fixed chemical potential, the critical hopping parameter $t_c=t_c(\Phi)$ indeed follows a fractal curve as the ``magnetic" flux is varied. As in the case of the superfluid-insulator transition in 2D rotating bosonic systems, this fractal phase boundary is directly related to the single-particle energy spectrum. \\

The author acknowledges P. Gaspard, P. de Buyl and A. Astudillo Fernandez for their support, and the F. R. S. - F. N. R. S. for financial support.



\begin{thebibliography}{99}
\bibitem{rmk} The result presented in Fig.\ref{hof_bord} doesn't depend on the supercell length $q$ which has been chosen arbitrarily among the prime numbers.
\end{thebibliography}
\end{document}